\documentclass[aps,prb,twocolumn,superscriptaddress]{revtex4}

\usepackage{amssymb}
\usepackage{amsfonts}
\usepackage{amsmath}
\usepackage{bm}
\usepackage{textcomp}
\usepackage{color}
\usepackage{longtable}
\usepackage{graphicx}
\usepackage{dcolumn}

\usepackage[a4paper=true,pagebackref=false]{hyperref}
\hypersetup{colorlinks,citecolor=blue,filecolor=blue,linkcolor=blue,urlcolor=blue}

\begin{document}


\title{Reconciling results of tunnelling experiments on (Ga,Mn)As}


\author{Tomasz Dietl}
\email{dietl@ifpan.edu.pl}
\affiliation{Institute of Physics, Polish Academy of Science, al.~Lotnik\'ow 32/46, PL-02-668 Warszawa, Poland}
\affiliation{Faculty of  Physics, University of Warsaw, PL-00-681 Warszawa, Poland}
\author{Dariusz Sztenkiel}
\affiliation{Institute of Physics, Polish Academy of Science, al.~Lotnik\'ow 32/46, PL-02-668 Warszawa, Poland}


\date{\today}

\begin{abstract}
 A theoretical model is presented which allows to reconcile findings of scanning tunnelling spectroscopy for (Ga,Mn)As [Richardella {\em et al.} Science 327, 66 (2010)] with results for tunneling across (Ga,Mn)As thin layers [Ohya {\em et al.}  Nature Phys. 7, 342 (2011);  Phys. Rev. Lett. 104, 167204 (2010)]. According to the proposed model, supported by a self-consistent solution of the Poisson and Schroedinger equations,  a nonmonotonic behaviour of differential tunnel conductance as a function of bias is associated with the appearance of two-dimensional hole subbands rather in the GaAs:Be electrode  than in the (Ga,Mn)As layer.

\end{abstract}


\maketitle

In a recent Article Ohya {\em et al.}\cite{Ohya:2011_NP} presented comprehensive investigations of tunnelling spectra obtained for high quality Au/(Ga,Mn)As/AlAs/GaAs:Be junctions. Similarly to previous studies of double barrier heterostructures by the same group,\cite{Ohya:2010_PRL} oscillations of differential conductance d$^2I/$d$V^2$ {\em vs.} $V$ for negative bias $V$ were interpreted in terms of resonant tunnelling involving Schottky barrier Au/(Ga,Mn)As and quantized subbands in the valence band of (Ga,Mn)As. By adjusting the corresponding parameters, positions of resonances in the studied samples could be explained provided that the valence band in (Ga,Mn)As (i) remains separated from the Mn impurity band; (ii) is weakly perturbed by disorder; (iii) its exchange splitting is very small even in samples with high Curie temperatures. While these findings corroborate conclusions of some optical studies,\cite{Burch:2006_PRL} they are striking, particularly in view of results of scanning tunnelling spectroscopy \cite{Richardella:2010_S}, which did not provide any evidence for the presence of an impurity band and pointed to a strong influence of disorder on the valence band of (Ga,Mn)As, especially comparing to the case of GaAs:Be. Thus, scanning tunnelling spectroscopy,\cite{Richardella:2010_S} supported the valence band conduction model \cite{Jungwirth:2007_PRB} of ferromagnetic (Ga,Mn)As.

\begin{figure}[!h]
\begin{center}
 \includegraphics[width = 8cm]{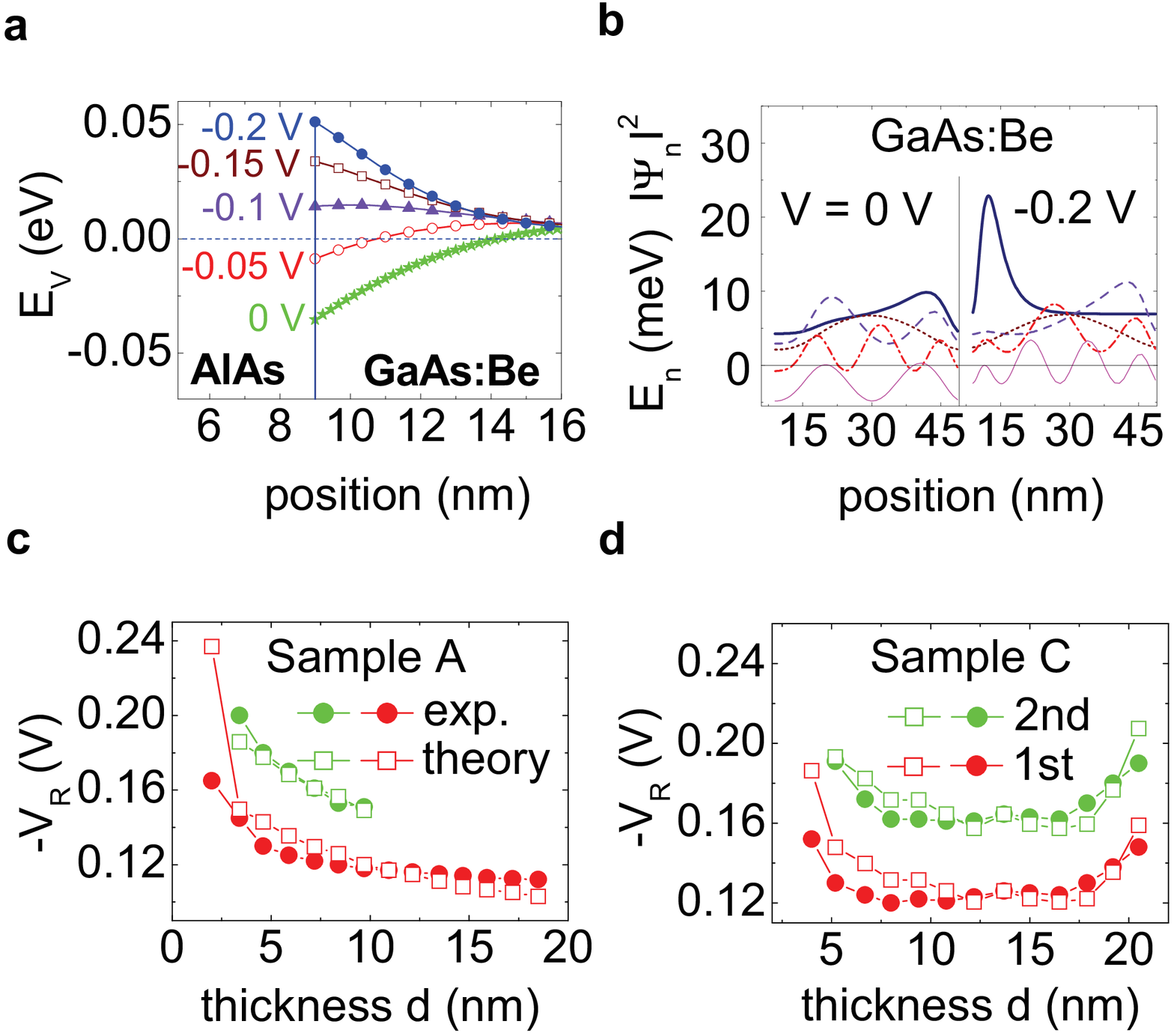}
\end{center}
\caption{ {\bf Formation and tunnelling {\em via} 2D interfacial subbands in GaAs:Be.} {\bf a}, Bias dependence of the valence band edge in GaAs:Be computed by solving self-consistently the Poisson and Schroedinger equations within 6 bands' $kp$ model assuming that GaAs:Be extends over 40 nm (for $1 \cdotp 10^{18}$ and $5 \cdotp 10^{20}$ holes per cm$^3$ in the valence band of GaAs:Be at V=0 and (Ga,Mn)As, respectively). {\bf b}, Corresponding energies and normalized wave function squares (zero values are at the level of corresponding eigenenergies) for $V = 0$ and  $-0.2$~V for five topmost hole states. The appearance of the 2D interfacial state becomes visible at $V < -0.1$~V.  {\bf c,d} Comparison of measured (full points, ref. \onlinecite{Ohya:2011_NP}) and calculated (empty points) values of resonant voltages $V_R(d)$ for 1$^{st}$ and 2$^{nd}$ tunnelling features as a function of (Ga,Mn)As thickness $d$.}
\end{figure}

We would like to propose here an alternative interpretation of the results,\cite{Ohya:2011_NP}  which makes it possible to reconcile the tunnelling data of refs.~\onlinecite{Ohya:2011_NP}  and \onlinecite{Richardella:2010_S}. We start our considerations by noting that the Fermi energy in (Ga,Mn)As is about 0.1~eV above the valence band top in GaAs. \cite{Ohno:2002_PE} This allows to fabricate magnetic tunnel junctions of (Ga,Mn)As/GaAs/(Ga,Mn)As, or even of (Ga,Mn)As/(Ga,In)As/(Ga,Mn)As.\cite{Elsen:2006_PRB} Accordingly, in the case of (Ga,Mn)As/AlAs/GaAs:Be heterostructures, owing to a modulation doping effect, an interfacial region in GaAs:Be is depleted of holes as shown in fig.~1. By applying a negative bias, a flat band condition is reached at $V_{\text{fb}}  \approx -0.1$~V, followed by hole accumulation in GaAs:Be. Due to a relatively low acceptor density and disorder, weakly broadened two-dimensional (2D) hole subbands are then formed in non-magnetic GaAs:Be. Their appearance results in a resonant-like tunnelling and, thus, in a nonmonotonic dependence of d$^2I/$d$V^2$. In fact,tunnelling involving holes accumulated at a p-GaAs/(Al,Ga)As interface was observed in a p-type single barrier junction. \cite{Hickmott::1992_PRB} At the same time, the resolution of separated 2D subbands in (Ga,Mn)As $via$ resonant tunnelling is hampered  by high hole relaxation rates in this alloy. \cite{Sawicki:2010_NP}

	The above considerations are supported by computations employing the $nextnano^3$ Poisson solver (see, ref.~\onlinecite{Sawicki:2010_NP}), pointing indeed, as presented in Fig.~1, to the formation of interfacial electric subband in GaAs:Be. Moreover, we note that the area resistance product $RA$, as measured for samples C and D (ref.~\onlinecite{Ohya:2011_NP}) as well as for sample A (ref.~\onlinecite{Ohya::2011_Comment}) as a function of the (Ga,Mn)As thickness $d$, is determined by Au/GaMnAs and GaMnAs/AlAs/GaAs:Be resistances in series. Since the subsequent etching steps change only the former, the resonance positions are given by $V_R(d) = V_R(d_m) \cdotp RA(d)/RA(d_m)$, where $d_m$ is an intermediate thickness. As shown in fig.1, the calculation explains the experimental variations $V_R(d)$. Thus, the proposed model allows reconciling the findings of refs.~\onlinecite{Ohya:2011_NP} and \onlinecite{Richardella:2010_S} as well as explaining why exchange splitting and a significant contribution of the impurity band were not observed in the tunnelling spectra. \cite{Ohya:2011_NP,Ohya:2010_PRL}

This work was supported by FunDMS Advanced Grant of European Research Council.

\end{document}